\documentclass[runningheads]{llncs}
\usepackage{graphicx}
\usepackage{comment}
\usepackage{amsmath,amssymb} % define this before the line  numbering.
\graphicspath{ {images/} } 
\usepackage{float}
\usepackage{color}

\begin{document}

\pagestyle{headings}
\mainmatter

\title{Anysize GAN: A solution to the image-warping problem} % Replace with your title

%\begin{comment}
\titlerunning{Anysize GAN: A solution to the image-warping problem} 
\authorrunning{Connah, Kendrick. David, Gillespie. Moi Hoon, Yap} 
\author{Connah, Kendrick. David, Gillespie. Moi Hoon, Yap}
\institute{Manchester Metropolitan University}
%\end{comment}
%******************

\begin{comment}
\titlerunning{Anysize GAN}
\author{Connah Kendrick\inst{1}\orcidID{0000-0002-3623-6598} \and
David Gillespie\inst{1}\orcidID{1111-2222-3333-4444} \and
Moi Hoon Yap\inst{1}\orcidID{0000-0001-7681-4287}}
%
\authorrunning{F. Author et al.}
\institute{Manchester Metropolitan University, Manchester, United Kingdom
\email{Connah.Kendrick@mmu.ac.uk}\\
\url{}\\}
\end{comment}  
\maketitle

\begin{abstract}  
	% intro   
	We propose an improvement to General Adversarial Networks (GANs) frameworks to resolve a inherent issue with deep learning, namely the detrimental pre-processing step, resizing. We develop a novel architecture that can be applied to existing latent vector-based GAN. Existing GANs for image generation requires images of uniform dimensions. However, publicly available datasets used commonly in pre-training and benchmarking, such as ImageNet, contain a multitude of resolutions. Resizing images to a uniform resolution causes deformations, changing the image data in a destructive process. Our network does not require this preprocessing step, allowing training over the full datasets without the requirement of resizing images. We make significant changes to the data loading techniques to enable training on multiple resolutions. We make two significant modifications to the generator by adjusting the network inputs and a dynamic resizing layer. Finally, we adjust the discriminator to use Global Average Pooling (GAP). These changes allow multi-resolution datasets to be trained on without any resizing. As a proof of concept, we validate our results on the ISIC 2019 skin lesion dataset. We demonstrate our method can successfully generate realistic images at different sizes without issue. Furthermore, the network demonstrates a full understanding of the spatial integrity of the generated images. We will release the source codes upon paper acceptance.
\end{abstract} 

\section{Introduction}

Generative Adversarial Networks \cite{Goodfellow2014} are a branch of deep learning, using two opposing neural networks to synthesize realistic images and determine real and fake images. Commonly used in multiple research areas, showing a drastic increase in the performance of deep learning and paved the way for better results, such as in skin lesion diagnosis \cite{Wu2019}, image inpainting \cite{pathak2018}, and object classification \cite{Katsilometes2018}. However, although these are separate tasks, the current solutions use data processing with deep learning. In which, a critical data factor is commonly overlooked, lesion shape and size is a vital part of diagnosing skin lesions; old paintings and digital art is rarely uniform; and deforming spatial data in object detection remove critical aspects. In each of these cases, changing the image shape will remove critical information that cannot be restored; thus resizing should be avoided.

\begin{figure*}
	\centering
	\includegraphics[scale=0.35]{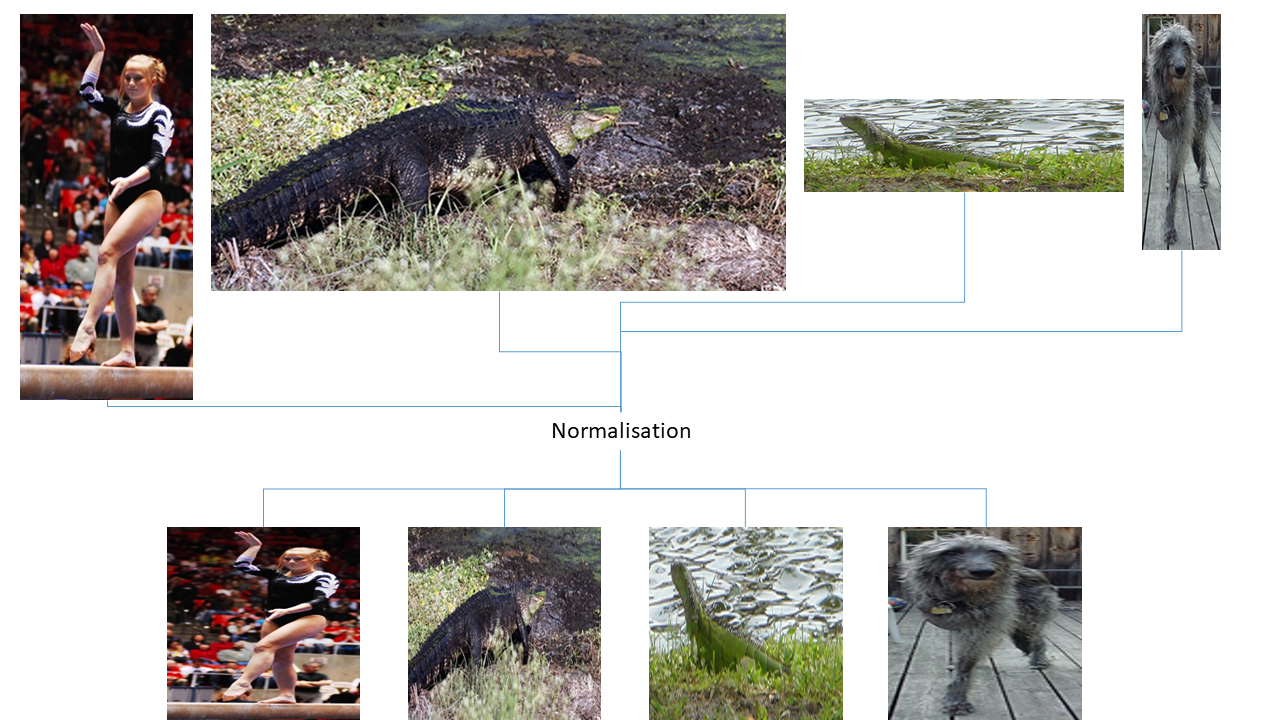} 
	\caption{An illustration of the resizing process that removes, duplicates and changes spatial data within an image.}
	\label{fig:Norm}
\end{figure*}

The state-of-the-art GANs require fixed size input images and generate fixed size output images. Requiring fix-sized images is restrictive as many natural images come in multiple resolutions. Furthermore, even in medical fields for the same task, such as skin lesion photography, X-rays and Magnetic Resonance Imaging (MRI), we cannot dictate that their equipment should meet our needs as resizing can remove the details used for diagnosis. In standard neural networks, it is common to use a square image, e.g. $224\times224$ in a 1:1 ratio. However, most cameras use alternative aspect ratios, such as 16:9 in television, which means almost half the horizontal data would be lost in resizing. The shape of the image can also be vital to the context, such as taking a picture of a car or a landscape we would use a horizontal image (16:9). Whereas, as ``selfie" or tower, we would use a vertical (9:16) image. By resizing the images, the shape of the object becomes distorted, as shown in Fig \ref{fig:Norm}. 
The original images have natural proportions, but when resizing, the images become ``dwarfed", key details, such as the angle of the hand seems more extreme. The distortions shown are an example of when images are down-sampled. In contrast, due to the high demand in using high resolution images, distortions from upsampling images become apparent. As a consequence of upsampling process from low resolution to high resolution, the images are often blurry and pixelated.

To illustrate the variation of image resolution, we summarise a few a popular benchmark datasets in computer vision research in Table \ref{tab:DatasetExplain}. The original format of ImageNet \cite{Russakovsky2015} consists of 773,565 images with 77,990 different image sizes. For ImageNet, there is a mean of 10 images per resolution, with a Standard Definition (SD) of 750. This shows a wide spread of image resolution in the dataset. Similar to ImageNet, CelebA \cite{Liu2015} contains various image resolution and only a small proportion are square images. However, this disparity is less obvious on images captured under controlled environment, such as Morph \cite{Karl2006}, with only 2 image resolution on portrait setting. But, in medical imaging, such as in ISIC \cite{Tschandl2018}, although it is under controlled environment, the variations in image resolution is caused by different brand of dermoscopic devices.

\begin{table} 
	\caption{A breakdown of the datasets and its image resolution. Mean$\pm$SD represents the mean and Standard Deviation (SD) of the number of images per resolution. Landscape refers to how many resolutions in the dataset have a greater width than height. Portrait refers to the number of resolutions that have a greater height than width. Square refers to a resolution that has equal width and height.}
	\begin{tabular}{| c | c | c | c | c | c | c |}
		\hline 
		Dataset  & Total Images & Resolution & Mean$\pm$SD & Landscape & Portrait  & Square\\ \hline   		
		ImageNet \cite{Russakovsky2015} & 773,565 & 71,990 & 10.75 $\pm$ 750.05 & 43,612 & 27,796 & 582 \\ \hline 
		CelebA \cite{Liu2015} & 202,599 & 62,091 & 3.26 $\pm$ 34.23 & 14,574 & 47,035 &482 \\ \hline
		Gwern Face \cite{Anonymous} & 302,623 & 648 & 467.01 $\pm$ 3223.24 & 362 & 271 & 15\\ \hline
		Morph \cite{Karl2006} & 55,134 & 2 & 27,567 $\pm$ 12,762 & 0 & 2 & 0 \\ \hline
		ISIC \cite{Tschandl2018} & 25,330 & 29 &  873.48 $\pm$ 2956.38 & 2 & 26 & 1 \\ \hline
	\end{tabular}  
	\label{tab:DatasetExplain}
\end{table}

The issue of resizing is thus two-fold. Firstly, resizing has been a necessity due to hardware limitations, but with the growth of cloud infrastructure and computational resources, this is less of an issue for some. Secondly, due to the fixed size nature of neural networks, such as no support of jagged matrices and the diversity of image resolutions in datasets resizing has become a common practice, overlooking the effect it has on the data that we trained. In this paper, we aim to provide a solution to overcome the issues caused by resizing.

To study on how resizing affects natural images, we use an example of CelebAHQ \cite{Liu2015} image. We use CelebAHQ to demonstrate as it used the progressive growing of GANs to generate the images, in factors of 2 ($128\times128$, $256\times256$, $512\times512$ and $1024\times1024$). In an ideal case, if we were to resize the images from $1024\times1024$  to $128\times128$, there should be no difference and the features of $1024\times1024$ would still be present in the resized images. However, as demonstrated in Table \ref{tab:resizeErr}, this is not the case. Resizing techniques create statistically different images where feature are distorted and changed, noting this is experimented on a square image. Although, the image may look the same visually, it has been shown by Goodfellow et al. \cite{Goodfellow2015}, even changes un-perceivable to humans may have an effect on neural network results.

\begin{table} 
	\caption{A demonstration of the destructive nature of resizing images, when a natural $1024\times1024$ image is downsampled to $128\times128$ and compared to what the image would look naturally. The results show the imperfect nature of resizing, there are severe structural and informational changes, even in naturally square images, such as the hat and hair are visually pixelated.The Diff map highlights the structural changes between natural image and a resized image.}
	\label{tab:resizeErr}
	\begin{tabular}{| c | c | c | c | c | c | c |}
		\hline 
		Natural $128\times128$ & Resized $128\times128$ & Diff map & Sampling & MSE & PSNR & SSIM \\ \hline   		
		\includegraphics[scale=0.5]{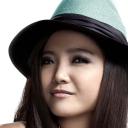}& \includegraphics[scale=0.5]{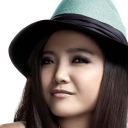}& 
		\includegraphics[scale=0.7]{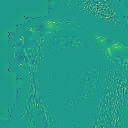}& Area & 18.6 & 35.4 & 0.97\\ \hline
		
		\includegraphics[scale=0.5]{"diff/00007"}& \includegraphics[scale=0.5]{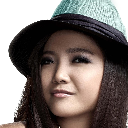}& \includegraphics[scale=0.7]{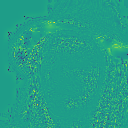}& Cubic & 166.3 & 25.9 & 0.84\\ \hline
		
		\includegraphics[scale=0.5]{"diff/00007"}& \includegraphics[scale=0.5]{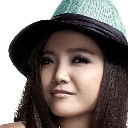}& 
		\includegraphics[scale=0.7]{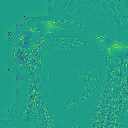}& Linear & 114.1& 27.6& 0.87\\ \hline
		
		\includegraphics[scale=0.5]{"diff/00007"}& \includegraphics[scale=0.5]{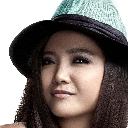}& \includegraphics[scale=0.7]{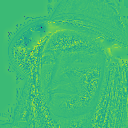}& Nearest & 316.3& 23.1& 0.78\\ \hline
	\end{tabular}  
\end{table}

\section{Related work} 
For some tasks, the networks are naturally more suited to generating images of differing sizes and require fewer adjustments, namely Fully Convolutional (FC) networks. FC networks are standard in tasks, such as inpainting \cite{pathak2018}, where the generator takes in an image. However, this requires adjustment to the network to support different sizes of images. In this paper, the most common type of GAN, i.e. latent vector-based GAN, will be discussed in this section. 

Latent vector-based GANs take a one dimensional (1D) fixed size vector as an input to the generator; this 1D vector is then passed through a series of Multi-Layered Perceptions (MLP). MLP layers are fixed size, as for each a learned weight is applied, causing much of the issues of GANs producing fixed sized images. Once passed the MLP in the generator convolutions are used, this stage is size invariant. The MLP layer is then once again used in the discriminator to determine if an image is real or fake, thus making the discriminator also size-dependent. 

This section will be split into two parts. Firstly, we present a review on size invariant issues. Secondly, due to there are no prior works on Anysize GAN, we review the related work in GANs. 

The issue of size invariant convolutional neural networks has been resolved, with two methods that provide a technique of taking an image of anysize and reducing the dimensions down to fixed size. He et al. \cite{He2014} introduced Spatial Pyramid Pooling (SPP), in which the input is fed through a multi-scale pooling into bins. The bins are then concatenated to create a fixed sized vector. Alternatively, Cui et al. \cite{Cui2017} implemented a simpler, but just as effective method, i.e. Global Average Pooling (GAP). GAP takes an average of each layer of the previous output, and because this is a fixed size based on the kernels in the convolutional layers, produces a fixed size vector. Both of these methods have shown improvements to the accuracy of neural networks in different fields \cite{He2014,Cui2017}. By implementing GAP or SPP for neural networks that take images as input, but use MLP layers, such as a discriminator it removes the fixed size limitation. 

As mentioned, one of the reasons for a natural reliance of resizing in this field comes from a lack of GPU power, where the focus was on small neural networks processing simple images, such as the MNIST and Fashion MNIST datasets \cite{lecun-mnisthandwrittendigit-2010,Xiao2017} which are small $28\times28$ images. Similar to Cifar10 and Cifar100 \cite{Krizhevsky}, the common image sizes are small ($32\times32$) to train deep learning. Many of the available datasets with a fixed size, such as these have either already gone through a preprocessing step \cite{zhou2017places}, or have been generated previously\cite{Liu2015}. Furthermore, with datasets, such as OpenImages \cite{OpenImages2}, the requirement for networks capable of processing a dynamic range of image sizes is needed. However, this does not remove the memory limitations. Thus, our network allows the preservation of image aspect ratio, during the downsampling process, preserving some spatial information.

One of the first research work in exploring deep neural networks for large scale image generation was by Radford et al. \cite{Radford2016} with DCGAN, they demonstrated the ability of neural networks to generate a large and high-quality image with a standard neural network. Their work has since been improved upon, in both terms of resolution and realism with StyleGAN and StyleGAN 2 \cite{Karras} among other neural networks. We focus on networks that dealt with size and resolution variations, such as Karres et al. \cite{Karras2017} in Progressive Growing of GANs (PGGAN). They used the trend for increasing resolution and developed a block based practice, which allows networks to extend to higher-resolution with minimal code changes. In which, this method is seen in many modern GANs such as StyleGAN. Karres et al. \cite{Karras2017} expanded the GAN structure by splitting the generator and discriminator into sub-blocks. Each block in the generator, is in essence, an up-sampling block, doubling the size of the input. Likewise, the block in the discriminator is down-sampling, reducing the size by half. By using the block theory when creating the model, the output scale can be increased and decreased easily. However, many works use transpose convolutions and up-sampling layers, restrict the output size by a factor of two, is not fully dynamic, unlike our solution. Another well-known GAN is SinGAN \cite{Shaham2019}, which can learn and produce realistic results with a single image, using pyramid convolutional structure, SinGAN demonstrates the ability to generate different sized images. However, SinGAN relies on input image and its FC, whereas, we adopt a more widely used latent vector style GANs, which is a much different challenge. Furthermore, SinGAN exceeds at learning the content of a single image and creating alternative representations of that image. However, it focuses on single images only matching content of that image i.e. on a dog the breed would always be the same, it cannot learn the differing breeds from a single image. In other words, the generated images are constraint by the context of the single trained image, and is unable to generate diverse features, e.g. for a skin lesion it will only produce samples of that image. Whereas, ours can produce unique images to expand datasets, like traditional GANs, such as DCGAN. 

\section{Methodology}

\begin{figure*}
	\centering
	\includegraphics[scale=0.36]{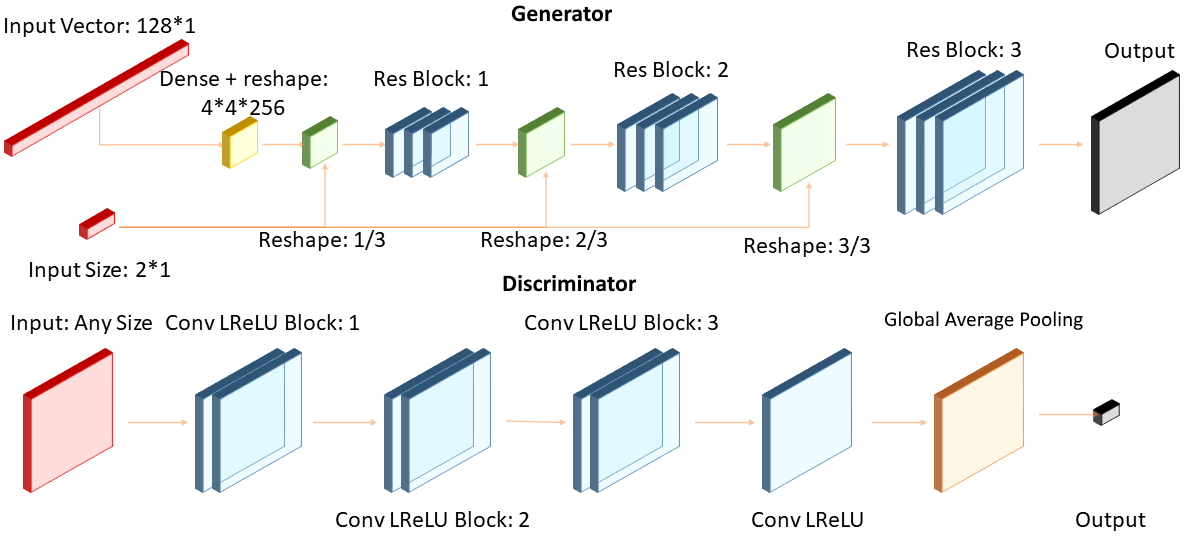}
	\caption{An illustration of the Anysize GAN network structure. Please note that for our network, we perform five resizes and ResNet blocks, with a ResNet block prior to the first reshape. The discriminator highlights the use of GAP.}
	\label{fig:Network}
\end{figure*}

To allow a GAN to generate variable sized images, we make three contributions. Firstly, we adjust the discriminator of the GAN to allow it to take images of varying sizes. GAP was used as it perform similar to SPP, but with a simpler implementation. Secondly, we adjust the generator to be multiple input. This allows the secondary input to dictate the size of generated images, as shown in Fig \ref{fig:Network}. Thirdly, design is the creation of a novel up-sampling layer. We used testing on both Nearest neighbour \cite{Olivier2012} shown in equation \ref{eq:nn} and bilinear interpolation in equation \ref{eq:Bilinear}. In which, bilinear showed improved results. In existing neural networks, the up-sampling layer works as a scalar, having values dictating how to multiply the size of the input in both height and width, so a value of 2 doubles the image size and is fixed once the model is compiled. Whereas our layer is dynamic, meaning at runtime, the variable of the resizing can be changed. Our layer takes the desired output size, allowing the layer to resize to the users requirement. This is a unique layer differing to how other GANs perform resizing. For our research, we tested if a single resize layer would work or a progressively growing technique. In which the progressive showed significantly better results from our empirical experiment. Hence, we use a progressive resizing over five layers, where the resizing happen before each convolutional block. During this stage, we pad all convolutions and remove other layers, such as max-pooling and upsampling that change the image size. 

\begin{equation}\label{eq:Bilinear}
f(x,y) \approx a_0 + a_1x + a_2y + a_3xy,
\end{equation}

\begin{equation}\label{eq:nn}
%\left.\begin{matrix}
%xRatio=\frac{W_{1}}{W_{2}}
%\\ 
%yRatio=\frac{H_{1}}{H_{2}}
%\end{matrix}\right\}
%W_{2},H_{2}\neq 0  
\begin{split}
xRatio=\frac{W_{1}}{W_{2}} , w_{2}\neq 0\\
yRatio=\frac{H_{1}}{H_{2}} , H_{2}\neq 0
\end{split}
\end{equation}

%Thirdly, we take a ResNet based GAN and edit the structure to remove spatial adjusting sections, such as pooling, up-sampling and strided convolutions. Additionally, we then pad all the convolutions; this means only our dynamic resizing layer adjustment has any control over the dimensions of the output image.

Additionally, we make adjustments to the loading and training loops. SinGAN, as its trained on a singular image, does not require a dataset. One of the known limitations of matrices, numpy, tensor and pytensor, is that each image in the matrix must be the same resolution, none jagged e.g. cifar10 is a 60,000 array of $32\times32$, but not image of $28\times28$. We adjust the whole training routine to work with different size matrices. During training, each batch in the epoch is a different size to the last. We loop the resolutions during training to allow better resolution generalisation and to prevent the network focusing on a single resolution. By implementing these steps into a single cohesive network, latent vector GANs gain the ability to generate images of any size on the fly. 

\section{Experimental Setup}

We implement Anysize GAN on ISIC 2019 dataset \cite{Tschandl2018}. As one of the critical features of skin lesion diagnoses is the spatial shape of the lesion, the resizing process should be avoided. Additionally, due to there is no benchmark algorithm in Anysize GAN, we compare the qualitative results with DCGAN \cite{Radford2016}. However, it is noted that our focus is to demonstrate the ability of Anysize GAN in providing a solution for the resizing problem, not on the accuracy.  

As highlighted in Table \ref{tab:DatasetExplain}, the ISIC dataset have variations in size. But due to the image size being greater then memory allowance, we group the images according to their aspect ratios. To maintain spatial data, we propose a method to group the images (without altering the aspect ratio). Let $I_{x_i,y_i}$ denotes an image from ISIC with width of $x$ and height of $y$, where $i$ represents the number of images with resolution $x \times y$. For each image resolution of $i$, $m_i$ denotes the maximum of $x_i$ and $y_i$, and $s_i$ represents the minimum of of $x_i$ and $y_i$, the images will be grouped, using equation \ref{eq:resize},

\begin{equation}\label{eq:resize}
m^\prime_{i}= 
\begin{cases}
M,& \text{if } m_i >= M\\
m_i,              & \text{otherwise}	
\end{cases}
\end{equation} 

where $m^\prime_{i}$ is the new larger side of the image, $M$ is the maximum size of the images (in our experiment, we set $M$ to 128, but it can be increased for researchers with high computational power). To calculate the corresponding size (the new smaller side), we maintain the ratio of the images by using equation \ref{eq:resize2}.

\begin{equation} 
\begin{aligned}\label{eq:resize2}
r = m_i/ m^\prime_{i}\\
s^\prime_{i}= s_i/r
\end{aligned}
\end{equation}

where $r$ is the ratio and $s^\prime_{i}$ is the new smaller side of the image. By using this method, we maintain the aspect ratios and preserve spatial information.

%machine Specs, batch size, ratio maintianace, loading and training the networks
The network was trained using Tensorflow \cite{Abadi2016} with Keras \cite{Chollet2016} as the foreground API. Due to memory limitations, we employed an image aspect ratio preserving resizing. Using equation \ref{eq:resize}, we resize to have the longest length equal to $128$. The maximum length of the images is to allow training on a batch size of 16, without causing memory issues. This means the network still train on different size images, but the unique resolutions will be less as some sizes will stack; for example, if we want to have a maximum size of $128\times128$, an image of $64\times128$ would not be resized, but an image of $128\times256$ will be resized to $64\times128$, reducing the number of resolutions in the dataset. For our experiment, we used a machine with an RTX 2080 Ti (11GB) GPU, 128GB RAM and an Intel i7-7820x CPU on Windows 10. We used python 3.6 with Tensorflow version 1.13.1 and Keras 2.2.4 to design and run the models. 

\section{Results}

As this is the first attempt in Anysize GAN, we compare the performance of our proposed method with DCGAN \cite{Radford2016} on the ISIC 2019 dataset. To enable fair comparison, we train DCGAN with input images of $128\times128$, similar to the maximum resolution ($M$) of our Anysize GAN. We train both networks for 180 epochs and report the results qualitatively and quantitatively.

%Connah, what do you mean?

%DCGAN allows more pixels to be seen over the dataset, but has less spatial data due to DCGAN removing the spatial resolution. Whereas, Anysize gets less information but preserves the spatial information of the lesions.

\begin{table}
	\centering
	\caption{Visual comparison of the generated images: Anysize GAN (left) and DCGAN (right).}
	\begin{tabular}{| c | c |}
		\hline 
		Anysize & DCGAN  \\ \hline   		  
		\includegraphics[scale=1]{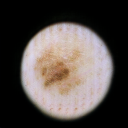} & \includegraphics[scale=1]{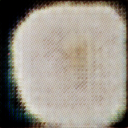} \\ \hline
		\includegraphics[scale=1]{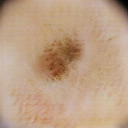} & \includegraphics[scale=1]{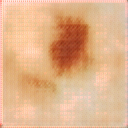}  \\ \hline
		\includegraphics[scale=1]{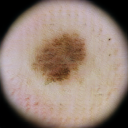} &  \includegraphics[scale=1]{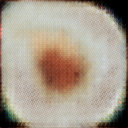}\\ \hline
	\end{tabular} 
	\label{tab:AnyVsDC}
\end{table}

As shown in Table \ref{tab:AnyVsDC}, the results from both networks are visually comparable. However, DCGAN has ``patchy" artefacts. In-comparison, Anysize GAN manages to create a smooth surface for both the mole and  the camera lens, without ``patchy" artefacts. It is widely known that quantitative measures on generated images (without ground truth) is a difficult task in GAN. To compare the performance of Anysize GAN and DCGAN quantitatively, we compare the generated datasets with the original ISIC dataset by using Inception Score (IS) \cite{Salimans}. As shown in Table \ref{tab:AnyVsDCScore}, we observe that Anysize GAN generates the dataset with diversity closer to the original ISIC dataset. This shows that Anysize GAN is comparable to other GANs of fixed size, even when the network is trained on varying sized images. 

\begin{table}
	\centering
	\caption{The inception score (IS)) of the Anysize GAN and DCGAN on 10,000 generated images and the original ISIC 2019 dataset.}
	\begin{tabular}{| c | c |}
		\hline 
		Images & IS \\ \hline  
		ISIC original & 4.3768 $\pm$ 0.2835 \\ \hline  		  
		Anysize GAN & 3.6063 $\pm$ 0.07010\\ \hline
		DCGAN & 2.6734 $\pm$ 0.04267\\ \hline
	\end{tabular} 
	\label{tab:AnyVsDCScore}
\end{table} 

To highlight the spatial awareness capable in the Anysize GAN, we demonstrate its capability in two ways.
\begin{itemize}
	\item Random latent vector and size: most GANs are demonstrated using a random latent vector, likewise we perform the same evaluation, but with random size. This shows our network is capable of generating a diverse amount of image not only in content but size. 
	\item Fixed latent vector, but alternative sizes: This is critical in demonstrating the affected network that generates alternative size images. This is because the resulting content of the latent vector should be almost identical, but as the image gets ``bigger" a zoom effect, such as more skin in the image should appear. This demonstrates the true spatial understanding of the network.
\end{itemize}

Table \ref{tab:outputs} shows an example of randomly generated skin lesion images using random latent vectors. This shows the network can produce results comparable to other state-of-the-art networks. In Table \ref{tab:outputsSameVector}, we demonstrate the ability of the network to generate the same image, but with different resolutions. Furthermore, it demonstrates spatial awareness by maintaining the shape of the lesions and not stretching as the resolution increases.

\begin{table} 
	\centering
	\caption{Images generated from our proposed Anysize GAN, with random latent vectors. These images demonstrate that Anysize GAN can generate realistic and diverse images, as well as different sizes.}
	\begin{tabular}{| c | c | c |}
		\hline 
		$128\times128$ & $89\times128$ & $84\times128$ \\ \hline   		  
		\includegraphics[scale=1]{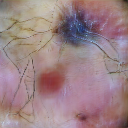} & \includegraphics[scale=1]{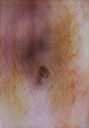} & \includegraphics[scale=1]{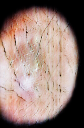}\\ \hline
		\includegraphics[scale=1]{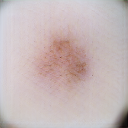} & \includegraphics[scale=1]{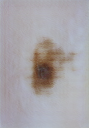} & \includegraphics[scale=1]{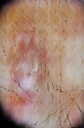}\\ \hline
		\includegraphics[scale=1]{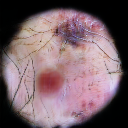} & \includegraphics[scale=1]{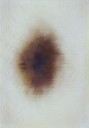} & \includegraphics[scale=1]{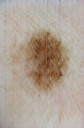}\\ \hline
	\end{tabular} 
	\label{tab:outputs}
\end{table}

\begin{table}
	\centering
	\caption{Images from Anysize GAN, with constant latent vector, but with different input size.}
	\begin{tabular}{| c | c | c | c | c |}
		\hline 
		size  & Training samples & Example 1 & Example 2 & Example 3 \\ \hline   		  
		84*128 & 153 &\includegraphics[scale=0.5]{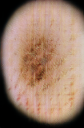} & \includegraphics[scale=0.5]{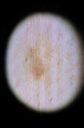} & \includegraphics[scale=0.5]{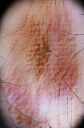}\\ \hline 
		
		85*128 & 1631 &\includegraphics[scale=0.5]{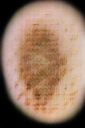} & \includegraphics[scale=0.5]{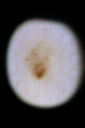} & \includegraphics[scale=0.5]{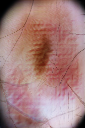}\\ \hline
		
		89*128 & 24 &\includegraphics[scale=0.5]{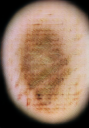} & \includegraphics[scale=0.5]{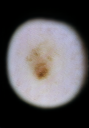} & \includegraphics[scale=0.5]{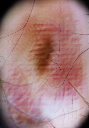}\\ \hline
		
		95*128 & 85 &\includegraphics[scale=0.5]{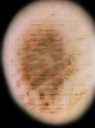} & \includegraphics[scale=0.5]{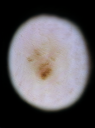} & \includegraphics[scale=0.5]{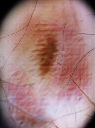}\\ \hline 
		
		96*128 & 10,899 &\includegraphics[scale=0.5]{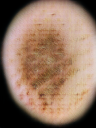} & \includegraphics[scale=0.5]{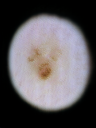} & \includegraphics[scale=0.5]{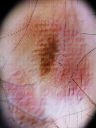}\\ \hline
		
		111*128 & 70 &\includegraphics[scale=0.5]{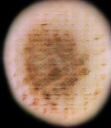} & \includegraphics[scale=0.5]{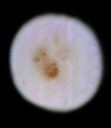} & \includegraphics[scale=0.5]{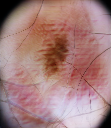}\\ \hline
		
		128*128 & 12,414 &\includegraphics[scale=0.5]{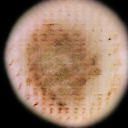} & \includegraphics[scale=0.5]{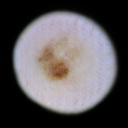} & \includegraphics[scale=0.5]{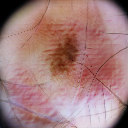}\\ \hline
		
	\end{tabular} 
	\label{tab:outputsSameVector}
\end{table}

\section{Conclusion} 

In this paper, we have highlighted the current restrictions of latent vector GANs:
\begin{itemize}
	\item Loss of spatial information due to resizing 
	\item Lack of diversity due to images of fixed size 
	\item Lack of general image understanding due to the warping of images
	\item Removes the natural image qualities from training
\end{itemize} 

To resolve these, we designed a new Anysize GAN with the following components:
\begin{itemize}
	\item A dynamic resizing layer 
	\item A multi-input GAN for size dictation
	\item Full end-to-end resolution impartial processing 
	\item An efficient anysize image matrix loading system.
\end{itemize}

We have shown a new way of designing neural networks that removes fixed size limitations. We show the adaptations allow the networks to be still comparable to other neural networks. This removes many of the issues with current datasets of multiple resolutions and aspect ratios needing to be resized. This work can potentially improve the use of deep learning for natural images processing. 

The work in this papers is a proof of concept to demonstrate the capabilities of neural networks in the generation of natural images by allowing multiple resolutions to be generated. Furthermore, future work can be performed: 
\begin{itemize}
	\item Resolution bias: the datasets have a natural bias to some resolution, research into how this affects the capabilities of the neural networks should be undertaken.
	\item Network backbone: we choose to use a common backbone for our network to aid the explainability of the network, but alternative backbone has potential to improve the quality of the results. 
\end{itemize}

\bibliographystyle{splncs}
\bibliography{references}

\end{document}